\documentclass[11pt,oneside,a4paper]{report}

\usepackage[english]{babel}
\usepackage[utf8]{inputenc}
\usepackage[T1]{fontenc}
\usepackage{type1ec}
\usepackage{graphicx}
\usepackage{amssymb}
\usepackage{pifont}
\usepackage{subfloat}
\usepackage{caption}
\usepackage{subfigure}
\usepackage{pdfpages}
\usepackage{gensymb}
\usepackage{appendix}
\usepackage{comment}
\usepackage{url}

\graphicspath{{./figuras/}}   

\begin{document}

\title{%
Mixtape Application:\\
\textbf{\Large{Last.fm Data Characterization}}}
\author
{\bfseries{Luciana Fujii Pontello \hspace{1cm} Pedro H. F. Holanda} \\
  \bfseries {Bruno Guilherme \hspace{1cm} Joao Paulo V. Cardoso}\\
  \bfseries {Olga Goussevskaia \hspace{1cm} Ana Paula C. da Silva}\\\\
  Computer Science Department\\ Universidade Federal de Minas Gerais (UFMG), Brazil\\\\
{\tt\small lucianafujii@dcc.ufmg.br, holanda@dcc.ufmg.br, brunoguilherme@dcc.ufmg.br}\\
{\tt\small jpcardoso@ufmg.br, olga@dcc.ufmg.br, ana.coutosilva@dcc.ufmg.br}
}
\def\authorname{Joao Paulo V. Cardoso, Luciana Fujii Pontello, Pedro H. F. Holanda, Bruno Guilherme, Olga Goussevskaia, Ana Paula Couto da Silva}

\date{}
\maketitle

\tableofcontents

\chapter{Introduction}

There is a never-ending debate about music and what happened to it after the dawn of the 
streaming age. A few very quick revolutions happened between the first appearance of
MP3 files and illegal downloads and the current scenario, where most artists have their
full catalog on at least one free streaming service, and many questions regarding
the consequences of that process are left without answers.

Back in the day, the only ways for people to listen to music were to turn on the radio
or, when choice was important, to buy large vinyl records that would eventually be stored 
somewhere in the house, waiting to be chosen to leave the comfort of their thick sleeves
for a spin on the turntable. These records had two sides, and required an inevitable pause
between the first and the second half of every record, since someone would have to go
there and flip the record. This single fact, for instance, had a strong influence on how
records were planned and how the sequence of songs was laid out.

Music albums were also a notable case of product tying, since they were a way for record 
companies to sell a large number of songs at once even though many customers only wanted
a couple songs off each album. This situation started to change with the release of
singles, which were cheaper, smaller vinyl records containing only a couple of songs,
and, a little later, by the introduction of portable cassette tape machines, which allowed
people to copy songs from different sources (radio, records or other tapes) to magnetic
tapes, which were cheap, small and practical.

This was the birth of the personal playlist. Tapes were the first time people could piece
together whatever tracks they wanted to listen to in whatever order they wanted and
then effectively listen to them, regardless of the popularity of the artists involved,
their likeliness to appear on the radio or the price of their LPs. The CD came as a more
practical alternative to that system, but, down deep, the album mentality was the same.

Of course, these are baby steps when compared to the flexibility and speed of the current
music industry. Many people don't even store music on their own computers anymore and,
in their cloud profiles, have access to music collections so large that they would
probably take up the space of a modern house in terms of physical records.
Entire albums can be bought, leased, shared or ``stolen'' in a few minutes over broadband
connection, and crafting and delivering a mixed tape to a friend only takes a few clicks.

This abundance of material creates a very urgent need for services that somehow make
it easier to find music that is interesting and relevant. There are many recommendation
services that try to address this problem in different ways. Some systems are able
make taste-based recommendations based on usage patterns, after a user has used them
for long enough. Other systems are based on curated lists of newly released and trending
items or artists, such as the user-made playlists on mixcloud\footnote{\url{www.mixcloud.com}}
or the professional dj-curated streams on 22tracks\footnote{\url{22tracks.com}}. 
Otherwise, the typically available navigation functions in media players and online
streaming services are mostly based on filtering by attributes, like title, artist
or genre, or return a list of similar items, computed using collaborative filtering techniques.

Now that we're witnessing the consolidation of these streaming services, it is more important
than ever to have a better understanding of how people listen to music. This is crucial
for the success of any music-related service, especially the ones that intend to somehow
recommend new music, or create sequences or playlists for different people to listen to.

One of the simplest ways to do that is to look at how people behave in the internet
- more specifically, on Online Social Networks (OSNs).
These websites are meant to extend the individual experience of society by bringing social
activities to the internet, allowing people to discuss their favorite subjects and share
their thoughts, opinions and feelings in a distributed and independent way, making them
an interesting source of many different kinds of information.

Some of these networks carry information that is particularly valuable because it does not
require active behavior to gather itself. This way, it reflects true habit
and behavior that would otherwise have to be observed in an intrusive way. A great
example of an ingenious way to collect such information is Last.Fm\footnote{\url{www.last.fm}}:

Last.Fm is an online social network for music fans. It has a very ingenious way of
operating: Users deliberately install a lightweight crawler on their personal computers
to keep track of what they are listening to. That is called \textit{scrobbling}.
Once users have their musical history up on the website, they can interact with
other users, make virtual ``friends'' (or find real ones that have profiles)
and add them to their own profiles, leave messages, navigate through other users'
profiles to see what they listen to, among other things. It even integrates with
streaming services such as Spotify\footnote{\url{www.spotify.com}}, allowing users
to upload their streaming histories directly to last.fm and listen to songs.

Even though the way people listen to music has changed drastically over the last few
years, there is still little research characterizing these phenomena.
In this work, we focus our attention on analyzing how people share their music
listening habits with other people via the internet. In particular, we analyze
\textit{last.fm}: We seek to characterize last.fm user behavior using our analysis to shed 
light on how users interact in this Online Social Network (OSN), how their
preferences and activities may affect and be affected by content popularity
dynamics and, especially, how they listen to music.

We seek to answer the following key questions:
\begin{itemize}
  \item What is the Last.fm user profile, age, gender, location?
  \item How is user activity distributed among content?
  \item How is the distribution of tags associated with songs?
  \item How can we use the knowledge of our data collection to improve the
    similarity measures we intend to derive from it?
\end{itemize}

Our analysis reveals interesting details about the operation of last.fm.
In Particular, we show that users are young, and most are located in the United States.
We also learn that there is a large difference between the amount of people
who listen to popular songs and obscure songs, and that popularity should be taken
into account to build concepts of similarity between songs, for recommendation purposes.
And finally, we notice that users explore the tag functionality of the network quite
heavily, and that important information can be derived from these tags and used for
similarity and recommendation purposes.

The report is structured as follows. We discuss related work about music similarity
and recommendation in chapter~\ref{sec:related}. In chapter~\ref{sec:dataset} we present
detailed information about the dataset used for this study. In chapter~\ref{sec:user}
we describe our detailed analysis of the user-related information in the dataset. In
chapter~\ref{sec:tracksandartists} we present our analysis of artists and tracks. In
chapter~\ref{sec:tags} we describe the analysis of tags. Finally, in chapter~\ref{sec:conclusions},
we present our conclusions.

\chapter{Related Work}\label{sec:related}

The most relevant aspects of recommendation systems are the premises they
are based on, and how they build upon them. One of the most common premises
is content similarity, and that it would be desirable to recommend content
that is somehow similar to content that a given user enjoys. So, We present
a small compilation of related works in the area of Media Similarity,
Music Recommendation and Playlist Generation.

\textbf{Media item-to-item similarity computation:} 
Several research studies offer technical solutions to define similarity between a given pair of items, and they divide themselves among two schools. Some studies develop similarity measures that are based solely on content (objective approach), being that audio or video information — for instance, the spectral or rhythmic content of songs.
On the other hand, other studies develop similarity measures that are based on user-generated
data and tags, also known as collaborative filtering (subjective approach). In the context of music, 
different approaches to define item-to-item similarity have been studied extensively, such as
content-based measures~\cite{mcfee2012learning, cano2005content, logan2001music,
flexer2014inter, aucouturier2005way, van2013deep},
that analyze spectral or rhythmic properties of songs, and user-based
measures~\cite{minsu2015, konstas2009social, shepitsen2008personalized,
goussevskaia2008web, farrahi2014impact, kuhn2010social}, that
analyze user listening habits in online social networks or user-generated
tags~\cite{shepitsen2008personalized}.
Many of these studies aim at building recommendation
systems~\cite{mcfee2012learning, cano2005content,
van2013deep, farrahi2014impact, turnbull2014using, shepitsen2008personalized}.


\textbf{Music playlist generation:}  Another related line of research to music
navigation systems is \textit{playlist generation}. Several people have
addressed this problem from different perspectives. There are techniques that
use statistical analysis of radio streams~\cite{MailletEtAl_2009,
turnbull2014using, chen2013multi, chen2012playlist}, are based on
multidimensional metric spaces~\cite{chen2012playlist, goussevskaia2008web,
moore2013taste, moore2012learning, kuhn2010social}, explore audio
content~\cite{logan2002content},
and user skipping behavior~\cite{pampalk2005dynamic}.

In particular, authors in~\cite{pampalk2005dynamic} create playlists
based on audio music similarity and skipping behavior, while authors in
\cite{fields2008} use network flow analysis to generate playlists from a
friendship graph of artists on MySpace. Maillet et al~\cite{MailletEtAl_2009},
in turn, present an approach to generating steerable playlists from tags linked
to songs played in professional radio station playlists, and Chen et al~\cite{chen2012playlist}
model playlists as Markov chains, which are generated through the Latent Markov
Embedding machine learning algorithm. The heuristics in~\cite{pampalk2005dynamic}
use acoustic similarity and present linear complexity to return the \textit{next}
item to the user, which is feasible to navigate in a collection of 2,500 items,
as the authors did, but does not scale to larger collections, which is not only
desirable but necessary for modern standards.

\chapter{Data collection} \label{sec:dataset}

Last.fm provides a public API for collecting
information about songs, artists, albums and tags that have been
contributed by millions of users.

The collection of Last.fm was comprised of 2 steps: First we collected 0.28\%
of all songs from the Last.fm service (around 100 thousand songs) and their top
fans, and then we continued collecting the top songs of these fans and their
friends. We collected the top-25 most listened to songs of each user.

Our dataset was collected from November,
2014 to July, 2015, and contains 372,899 users (with their respective top-25
song lists), 2,060,173 songs, and 374,402 artists. Moreover, we also collected a
total of 1,006,236 user-generated tags, associated with songs. In particular, 47\% of songs have
had at least one associated tag in our dataset.

From a total of 2,060,173 songs in our database, 983,010 have MusicBrainz
Identifiers (MBID)\footnote{MBID is a reliable and unambiguous form of music
identification (\url{musicbrainz.org}).}. This is an important source of
reliability, since these entries are guaranteed to be unique, which eliminates
duplicate problems, and refer to the correct media items, which avoids
the occasional mismatch between the name of a song or artist in a user's
computer and the actual correct name of the item, not to mention non-released
live versions, non-official covers and so on.

\chapter{User profile}\label{sec:user}

We collected a total of 372,899 users from 237 countries. Table~\ref{tab:topUsersCountry}
shows the countries with most users we have collected.

\begin{table}[th]
\centering
  \begin{tabular}{lc}
  \hline
    Country & \# of users \\
  \hline
  US & 54248 \\
	BR & 31230 \\
	PL & 25882 \\
	RU & 22979 \\
	UK & 22515 \\
	DE & 18314 \\
	NL & 7237 \\
	CA & 6931 \\
	FI & 6356 \\
	FR & 6139 \\
  \hline
  \end{tabular}
  \caption{Top countries with most users}
\label{tab:topUsersCountry}
\end{table}

Many users did not declare their gender.  Of those who did, 201,096 said they
were male and 108,034 said they were female. Table~\ref{tab:genderPlaycount}
shows the number of average playcounts per gender. We can see that the average
number of playcounts of male users is bigger than that of female users. Users
that have not declared their gender have the smallest average playcount number.

\begin{table}
\centering
  \begin{tabular}{lc}
  \hline
    Gender & Average playcounts \\
  \hline
	Female & 82907.0770 \\
	Male & 113573.4874 \\
	Not informed & 56797.2135 \\	
  \hline
  \end{tabular}
  \caption{Average playcount per gender}
\label{tab:genderPlaycount}
\end{table}

\begin{figure}[t]
\begin{center}
\includegraphics[width=0.7\columnwidth]{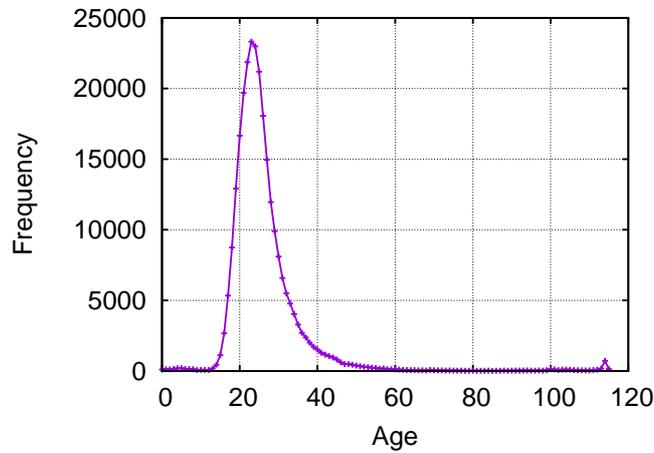}
\caption{Last.fm: Users' age distribution.}
\label{fig:userage}
\end{center}
\end{figure}

The age range reported by the users ranges from 0 to 115 years.
Figure~\ref{fig:userage} shows the distribution of users' ages.  We can see
that the great majority of Last.fm users are from 18 to 30 years old. Over
113,000 users did not report their age.

\begin{figure}[t]
\begin{center}
\includegraphics[width=0.7\columnwidth]{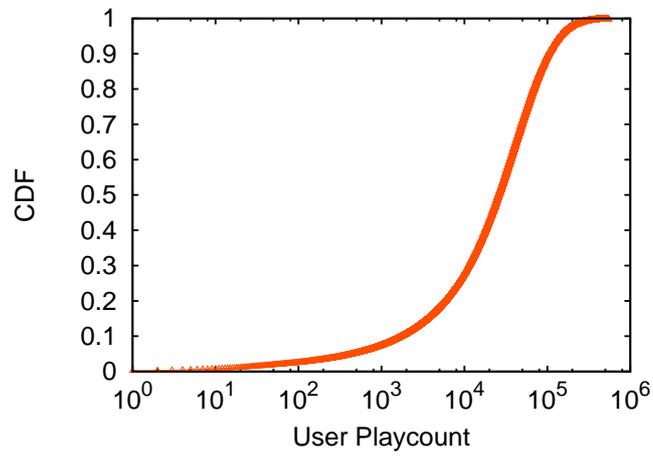}
\caption{Last.fm: CDF of users' playcounts.}
\label{fig:userplaycount}
\end{center}
\end{figure}

The Figure~\ref{fig:userplaycount} shows that approximately 27\% of the users
have listened to less than 10,000 songs, whereas 62\% of the users have
listened to 10,000 to 100,000 songs. In general users listen to a large amount
of songs, making a vast listening history, showing that their engagement with Last.fm
service lasts for a significant period of time.

\begin{figure}[t]
\begin{center}
\includegraphics[width=0.7\columnwidth]{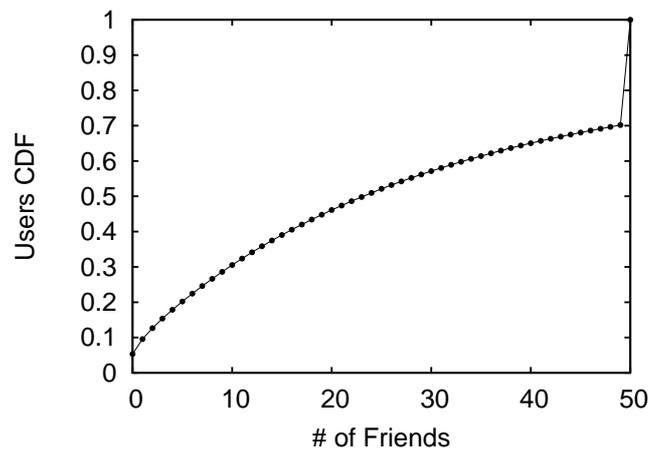}
\caption{Last.fm: CDF of users' friends.}
\label{fig:userfriends}
\end{center}
\end{figure}

Figure~\ref{fig:userfriends} shows the CDF of users' friends. Approximately 5\%
of the users have zero friends in Last.fm, and 50\% of the users have at most
25 friends. We have only collected a maximum of 50 friends of each user, explaining
the peak in the 50 number of friends point in the plot. Approximately 30\% of
the users have 50 friends or more.

\chapter{Songs and Artists}\label{sec:tracksandartists}

Sometimes, Last.fm registers different items which are actually the same song,
written in different ways. So, For this part of the analysis, we only took
into account the songs we could match with Musicbrainz IDs, in order to avoid
the risk of considering multiple issues related to track or artist name mismatch or repetition. 

\begin{figure}[t]
\begin{minipage}{.45\linewidth}
\centering
\includegraphics[width=\columnwidth]{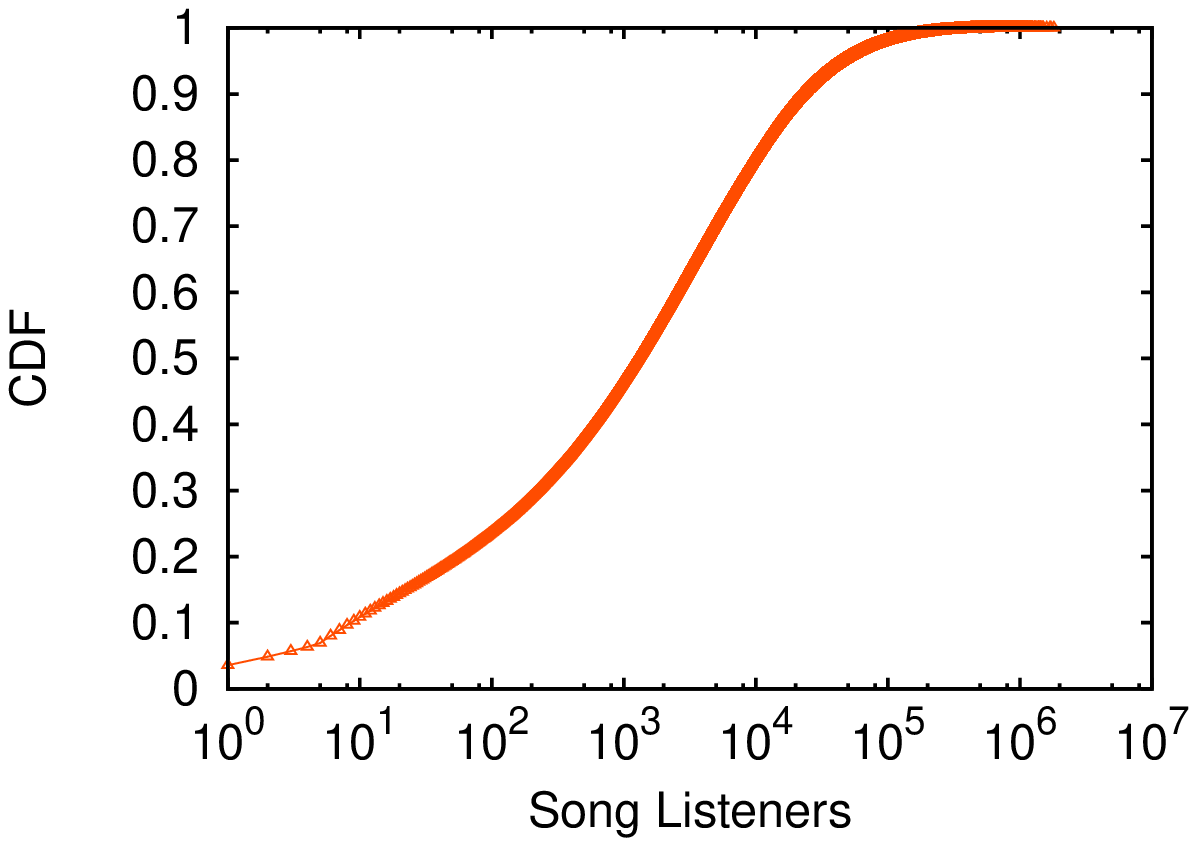}
\caption{Last.fm: CDF of song popularity by listeners.}
\label{fig:poptracklisteners}
\end{minipage}
\hfill
\begin{minipage}{.45\linewidth}
\centering
\includegraphics[width=\columnwidth]{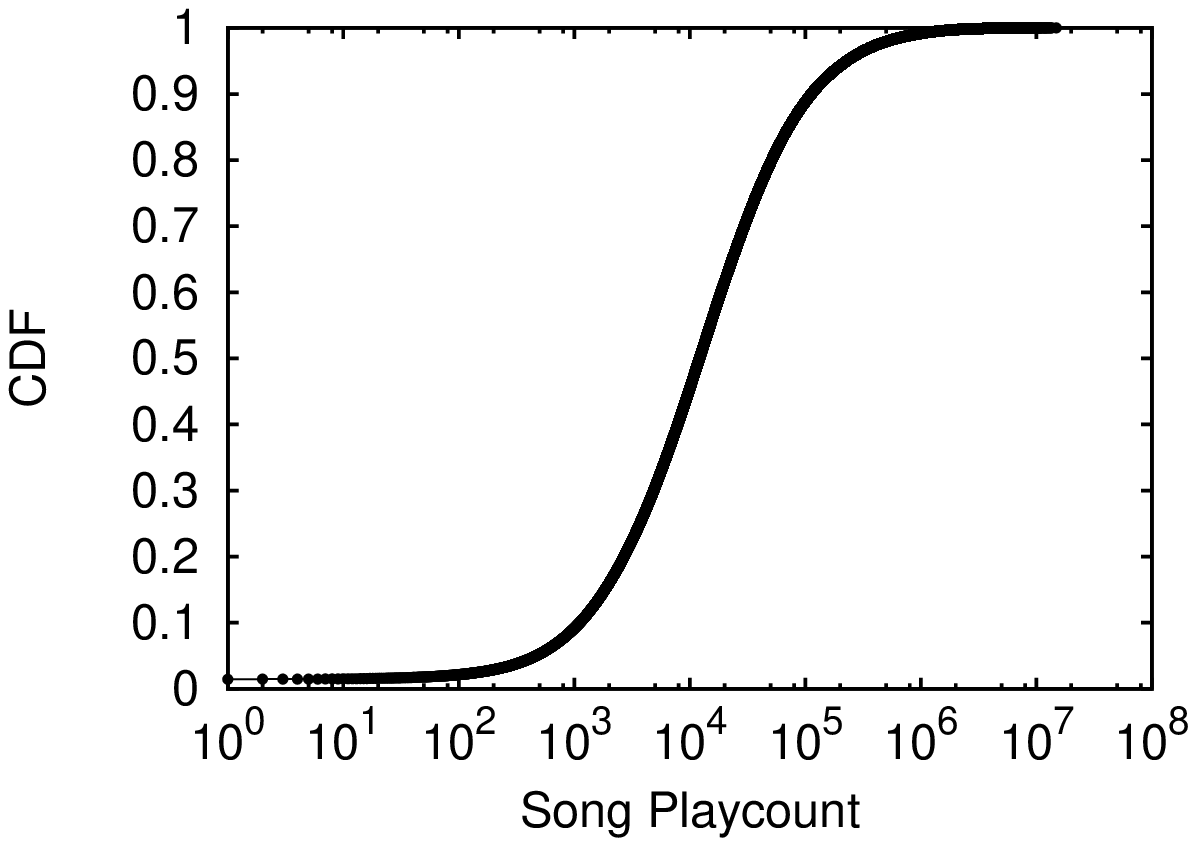}
\caption{Last.fm: CDF of song popularity by playcount.}
\label{fig:poptrackplaycount}
\end{minipage}
\end{figure}

Figures \ref{fig:poptracklisteners} and \ref{fig:poptrackplaycount} show the Cumulative
Density Function (CDF) of the popularity of songs by listeners and playcount.
We can observe that, in our dataset, 50\% of songs were listened to by at most
1,280 unique users (0.3\%); 10\%, in turn, attracted attention of more than 22,685
unique users (5.9\%). The most popular songs (top 1\%) were listened to by more than
155,468 users (41\%). The top 10 songs by number of listeners can be seen in
Table~\ref{table:songs}, each of them with more than 1.4M listeners.

In Figure~\ref{fig:poptrackplaycount} we can see that 10\% of the tracks have
1,000 playcounts or less and more than 50\% have more than 10,000 playcounts.

\begin{table}[h]
  \centering
\footnotesize
  \begin{tabular}{lll}
  \hline
  Track name & Artist & Number of listeners \\
  \hline
  Smells Like Teen Spirit & Nirvana         & 1,806,180 \\
  Mr. Brightside           & The Killers     & 1,716,969 \\
  Wonderwall              & Oasis           & 1,685,703 \\
  Come as You Are         & Nirvana         & 1,597,611 \\
  Clocks                  & Coldplay        & 1,507,981 \\
  Somebody Told Me        & The Killers     & 1,490,787 \\
  Take Me Out             & Franz Ferdinand & 1,462,621 \\
  Karma Police            & Radiohead       & 1,431,055 \\
  Viva la Vida            & Coldplay        & 1,431,034 \\
  The Scientist           & Coldplay        & 1,404,877 \\
   \hline
  \end{tabular}
  \caption{Top 10 songs}\label{table:songs}
\end{table}

\begin{figure}[t]
\begin{minipage}{.45\linewidth}
\centering
\includegraphics[width=\columnwidth]{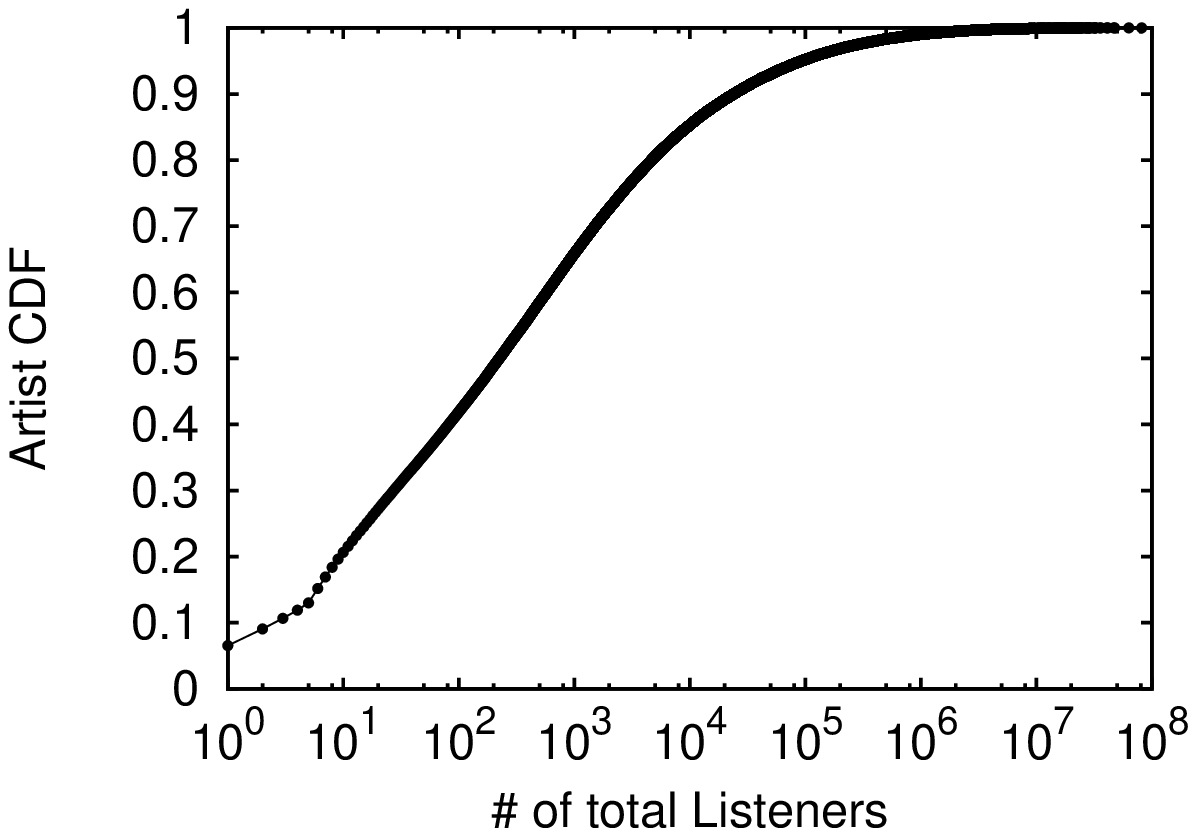}
\caption{Last.fm: CDF of artist popularity by number of total listeners.}
\label{fig:popartistlist}
\end{minipage}
\hfill
\begin{minipage}{.45\linewidth}
\centering
\includegraphics[width=\columnwidth]{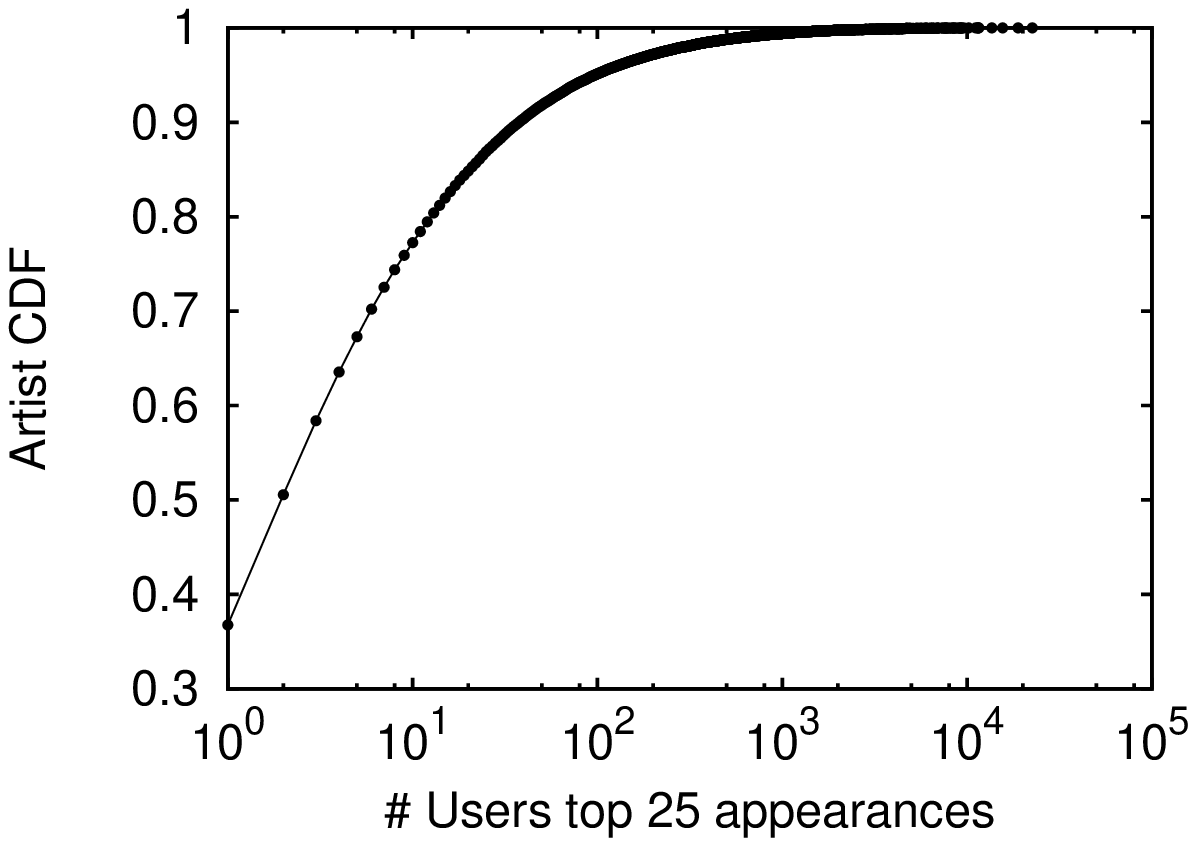}
\caption{Last.fm: CDF of artist popularity by users' top 25.}
\label{fig:popartist}
\end{minipage}
\end{figure}

Figure~\ref{fig:popartistlist} shows the Cumulative Density Function (CDF) of
the popularity of artists by their total listeners, retrieved from Last.fm.  The top 1\%
most popular artists ($\approx$ 3,577) were listened to by more than 764,000
users, while approximately 20\% of the artists were listened to by only 10 users.

Figure~\ref{fig:popartist} shows the Artist CDF by the number of users that
contain that artist in their top 25, in the users' history collected by us.
More than 35\% of the artists appear in only one user's top 25 songs, while 1\%
appear in more than 600 user's top 25 history.

\begin{figure}[t]
\begin{minipage}{.45\linewidth}
\centering
\includegraphics[width=\columnwidth]{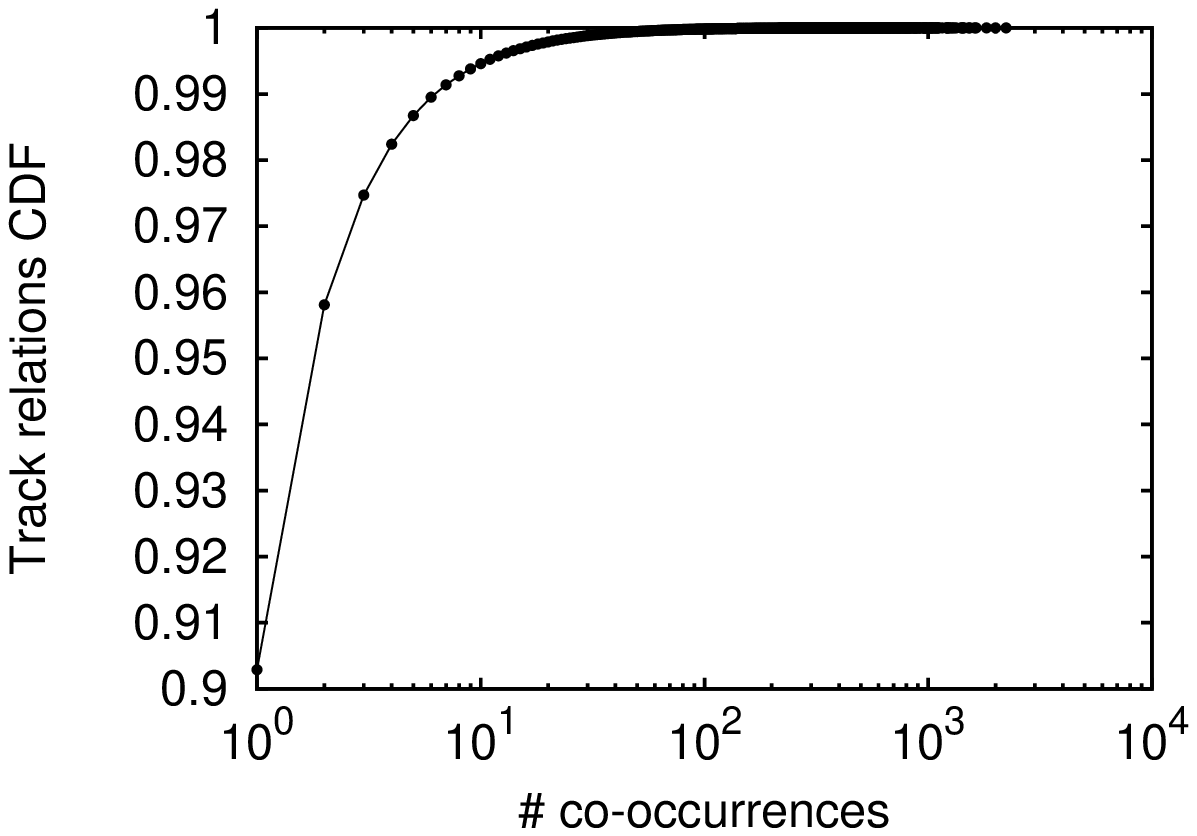}
\caption{Last.fm: CDF of song co-occurrence.}
\label{fig:coocc_cdf}
\end{minipage}
\hfill
\begin{minipage}{.45\linewidth}
\centering
\includegraphics[width=\columnwidth]{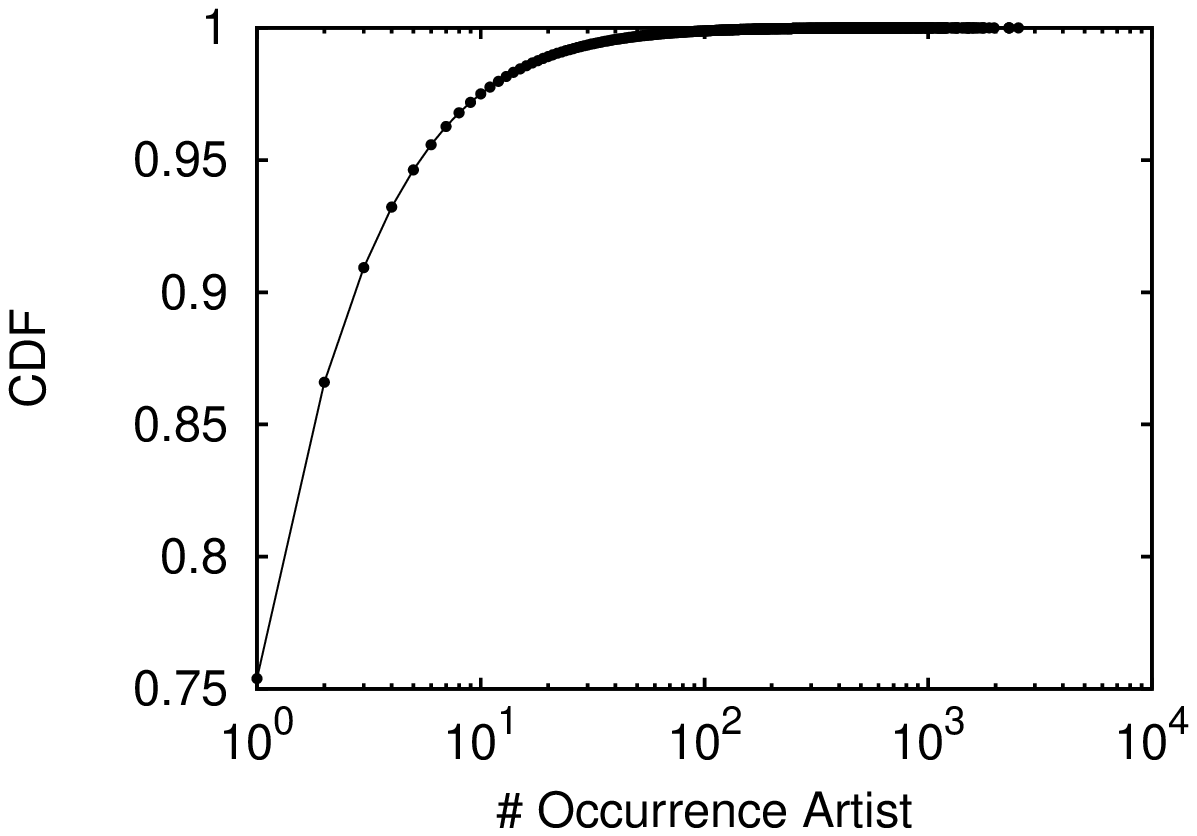}
\caption{Last.fm: CDF of artist co-occurrence.}
\label{fig:coocc_artist}
\end{minipage}
\end{figure}

Figure~\ref{fig:coocc_cdf} shows the co-occurrence CDF between songs that co-occurred at least
once. We define as a co-occurring relationship only the pair of songs that
appear in at least one user's top 25 listened songs.  We can see that more than
90\% of the co-occurrences happen in only one user listening history, while
only 1\% of the relationships co-occur 10 or more times.
If we consider all the NxN pairs of all songs, most of the songs don't even
co-occur in any user's top 25 songs. Therefore, initially, we can only infer
the similarity between a small percentage of the songs we have collected, and have
to infer similarity of the other songs from those similarities.

Figure~\ref{fig:coocc_artist} shows the co-occurrence CDF between artists
that co-occurred at least once. We define as a co-occurring relationship only
the pair of artists that appear in at least one user top 25 listened songs. We
can see that 75\% of the artists co-occur only in one user's top 25 songs, whereas
approximately 1\% of the artists co-occur in 22 users' top 25.

\chapter{Tags}\label{sec:tags}

Last.fm allows users to create and associate tags with the songs they listen
to. One tag can be associated to one song from one to 100 times,
allowing the users to show their agreement of a certain tag association
by adding to it. We collected a total of 1,006,236 user-generated tags,
associated with songs. 47\% of the songs have had at least one associated
tag in our dataset and considering only the songs with MusicBrainz ID,
75\% of the songs were associated with at least one tag.

The top 5 most popular tags (rock, alternative, pop, indie and electronic) were
associated with 541,527 songs. Table~\ref{table:tags} shows the most popular
tags and the number of songs associated with them.

\begin{table}[h]
  \centering
\footnotesize
  \begin{tabular}{ll}
  \hline
  Tag name & Tag count \\
  \hline
  rock & 167,610 \\
  alternative & 101,061 \\
  pop & 96,654 \\
  indie & 88,786 \\
  electronic & 87,416 \\
  alternative rock & 56,643 \\
  favorites & 56,508 \\
  beautiful & 51,870 \\
  love & 50,918 \\
  awesome & 42,364 \\
   \hline
  \end{tabular}
  \caption{Top 10 tags}\label{table:tags}
\end{table}

\begin{figure}[t]
\begin{center}
\includegraphics[width=0.7\columnwidth]{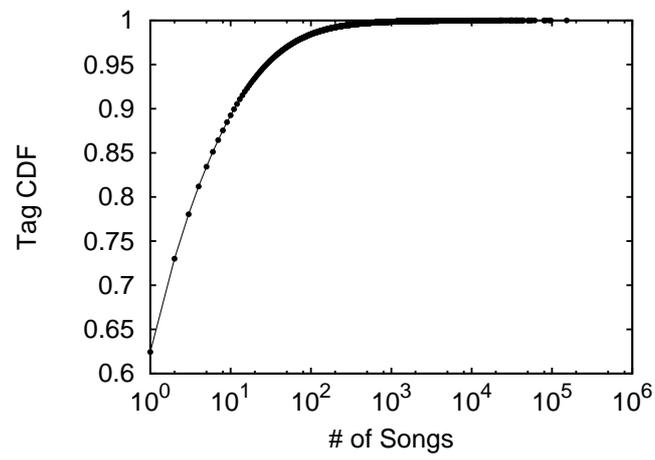}
\caption{Last.fm: CDF of tag popularity.}
\label{fig:poptag}
\end{center}
\end{figure}

Figure \ref{fig:poptag} shows the Cumulative Density Function (CDF) of tag
popularity using the number of songs associated with the tag: 62\% of tags have
been associated with only one song. In contrast, the top 5 most popular tags
were associated with more than 87,000 songs each.

\chapter{Conclusions}\label{sec:conclusions}

In this technical report we presented a broad characterization of Last.fm, an
Online Social Network (OSN) for music fans. From that characterization we
understood more about the profile of Last.fm users and how they listen to music
and interact with the OSN. We also collected insights about the data we have
collected, which is a small fraction of the information available in Last.fm
database.

We learned that the majority of the users we have collected are young, from 18
to 30 years old, and from various countries, but the majority of users are still
in the US. This will probably impact our similarity measures, which will be
more accurate for the songs that appeal to this young audience and that are
popular in the US, because we will have more co-occurrence data about these
songs. If we want to have better similarity metrics for a specific country or
region, it is fair to assume that it will be beneficial to collect more data
from users from that region as a priority.

We discovered that there are songs that are listened by only a handful of users,
while the top songs are listened by thousands users. This means that we will
probably achieve better similarity measures for the most popular songs. Based on
that information, we could develop new methods to gather more data about the
less popular songs. For instance, we could collect those songs top fans
histories, in order to gather more co-occurrence data from these songs. Or, in
the future, we could try to ally our similarity measures based on social data to
other techniques, for example, techniques that use the analysis of the content
of the songs, to derive similarity measures for new or unknown songs we want to
recommend.

Users also heavily used tags to classify the songs they listen to. From tags we
learned that the most common genre of song in Last.fm is rock, the tag
associated with the most songs in the social network. Since tags are present in
most of the songs that contain a MusicBrainz ID, we can take advantage of tags
to improve our similarity metrics, for instance, using tags to derive
similarity of songs with zero co-occurrences to other songs. Or we can develop
navigation techniques that take advantage of tags. Finally, we can even use
tags to evaluate our similarity metrics and Euclidean space, by analyzing if
songs with a given tag tend to be close together or spread around the map.

\bibliographystyle{plain}
\bibliography{relatorioLG}

\end{document}